\documentclass{elsart}
\usepackage{latexsym}
\usepackage{epsfig}
\usepackage{natbib}

\def\url#1{#1} 

\begin{document}

\begin{frontmatter}

\title{A Toy Model for Magnetic Extraction of Energy from Black
Hole Accretion Disc}

\author {DING-XIONG WANG$^{1}$, YONG-CHUN YE, AND REN-YI
MA}
\address{  Department of Physics, Huazhong University of Science
and Technology, Wuhan,430074,People's Republic of China
\thanksref{email} }

      \thanks[email]{E-mail: dxwang@hust.edu.cn}

\begin{abstract}

A toy model for magnetic extraction of energy from black hole (BH)
accretion disk is discussed by considering the restriction of the
screw instability to the magnetic field configuration. Three
mechanisms of extracting energy magnetically are involved.
(\ref{eq1}) The Blandford-Znajek (BZ) process is related to the
open magnetic field lines connecting the BH with the astrophysical
load; (\ref{eq2}) the magnetic coupling (MC) process is related to
the closed magnetic field lines connecting the BH with its
surrounding disk; and (\ref{eq3}) a new scenario (henceforth the
DL process) for extracting rotational energy from the disk is
related to the open field lines connecting the disk with the
astrophysical load. The expressions for the electromagnetic powers
and torques are derived by using the equivalent circuits
corresponding to the above energy mechanisms. It turns out that
the DL power is comparable with the BZ and MC powers as the BH
spin approaches unity. The radiation from a quasi-steady thin disk
is discussed in detail by applying the conservation laws of mass,
energy and angular momentum to the regions corresponding to the MC
and DL processes. In addition, the poloidal currents and the
current densities in BH magnetosphere are calculated by using the
equivalent circuits.

\end{abstract}


\end{frontmatter}


\section{INTRODUCTION}

\quad\quad As is well known, the Blandford-Znajek (BZ) mechanism
has been regarded as a reasonable process for powering the radio
jets in AGNs (Blandford {\&} Znajek 1977; Rees 1984). Recently the
BZ mechanism has been used as a central engine for powering
gamma-ray bursts (GRBs), where rotating energy of a stellar black
hole (BH) with magnetic field of $10^{15}gauss$ is extracted along
the magnetic field lines supported by a magnetized accretion disk
(Lee et al. 2000; Wang et al. 2002a). According to the BZ theory
the power extracted from the BH arises from the rotation of the
open magnetic field lines relative to the BH horizon, resulting in
the electromotive force (EMF) in an equivalent circuit (MacDonald
and Thorne 1982, hereafter MT82). The BZ power can be effectively
regarded as the power of the current dissipated on the
astrophysical load.

\quad\quad Recently much attention has been paid on the magnetic
coupling (MC) process, where energy and angular momentum are
transferred from a fast-rotating BH to its surrounding disk by
virtue of the closed field lines connecting them (Blandford 1999;
Li 2002a, hereafter L02; Wang et al. 2002b, 2003a, hereafter W02,
W03a, respectively).

\quad\quad In W02 we worked out an equivalent circuit to calculate
the BZ and MC powers. Very recently we discussed the condition for
the coexistence of the BZ and MC processes (CEBZMC), and found
that the state of CEBZMC always accompanies the screw instability
of the magnetic field connecting a rotating BH with its
surrounding disk (Wang et al. 2003b, 2004, hereafter W03b and W04,
respectively). It turns out that the screw instability will occur
at some place far away from the inner edge of the disk, if the BH
spin and the power-law index for the variation of the magnetic
field are greater than some critical values.

\quad\quad In this paper a new scenario for extracting rotational
energy of the disk matter is proposed by considering the
configuration of the magnetic field restricted by the screw
instability. To facilitate description, this mechanism is referred
to as the DL process, implying that energy and angular momentum
are extracted magnetically from \textbf{disk} to \textbf{ load}.
By using another equivalent circuit we derived the expression for
the DL power and torque under some assumptions on the unknown
astrophysical load.

\quad\quad This paper is organized as follows. In \S 2 the
restriction of the screw instability to the configuration of the
magnetic field is discussed, in which the three energy mechanisms
are contained. In \S 3 the expressions for the powers and torques
of the three mechanisms are derived in the two kinds of equivalent
circuits. We compare these powers with the variation of the two
parameters, i.e., the BH spin and the power - law index of the
magnetic field on the disk. It turns out that the DL power is
generally less than the BZ and MC powers, and it is comparable
with the latter two when the BH spin approaches unity. In \S 4 the
radiation from a quasi - steady thin disk is discussed in detail
by applying the conservation laws of mass, energy and angular
momentum to the regions corresponding to the MC and DL processes.
In \S 5 the poloidal current densities flowing from the BH
magnetosphere into the horizon and disk are calculated in the
regions corresponding to the three energy mechanisms. Finally, in
\S 6, we summarize our main results and argue that this model can
be used to extract clean energy from a rotating BH for powering
GRBs.

\quad\quad Throughout this paper the geometric units $G = c = 1$
are used, and the assumptions for BH accretion disk and the
magnetic field are adopted as given in W03b and W04.


\section{RESTRICTION OF SCREW INSTABILITY TO CONFIGURATION OF
MAGNETIC FIELD}

\quad\quad It is well known that the magnetic field configurations
with both poloidal and toroidal components can be screw instable
(Kadomtsev 1966; Bateman 1978). According to the Kruskal-Shafranov
criterion (Kadomtsev 1966), the screw instability will occur, if
the toroidal magnetic field becomes so strong that the magnetic
field line turns around itself about once. Recently some authors
discussed the screw instability with different conditions in the
BH magnetosphere (Gruzinov 1999; Li 2000a; Tomimatsu et el. 2001).
In W04 we discussed the screw instability of the magnetic field in
the MC process, and argued that the instability will occur if the
following criterion is satisfied,


\begin{equation}
\label{eq1} {\left( {{2\pi \varpi _{_{D}} } \mathord{\left/
{\vphantom {{2\pi \varpi _{_{D}} } L}} \right.
\kern-\nulldelimiterspace} L} \right)B_D^p } \mathord{\left/
{\vphantom {{\left( {{2\pi \varpi _{_{D}} } \mathord{\left/
{\vphantom {{2\pi \varpi _{_{D}} } L}} \right.
\kern-\nulldelimiterspace} L} \right)B_D^p } {B_D^T }}} \right.
\kern-\nulldelimiterspace} {B_D^T } < 1,
\end{equation}

\noindent where $L$ is the poloidal length of the closed field
line connecting the BH with the disk, $B_D^p $ and $B_D^T $ are
the poloidal and toroidal components of the magnetic field on the
disk, respectively, and $\varpi _{_{D}} $ is the cylindrical
radius on the disk and reads


\begin{equation}
\label{eq2} \varpi _{_{D}} = \xi M\chi _{ms}^2 \sqrt {1 + a_ * ^2
\xi ^{ - 2}\chi _{ms}^{ - 4} + 2a_ * ^2 \xi ^{ - 3}\chi _{ms}^{ -
6} } .
\end{equation}

In equation (\ref{eq2}) $M$ and $a_ * $ are the BH mass and spin,
respectively. The parameter $\xi \equiv r \mathord{\left/
{\vphantom {r {r_{ms} }}} \right. \kern-\nulldelimiterspace}
{r_{ms} }$ is the radial coordinate on the disk, which is defined
in terms of the radius of the marginally stable orbit, and $r_{ms}
$ is related to $M$ and $a_ * $ by the following relation (Novikov
{\&} Thorne 1973),


\begin{equation}
\label{eq3} \left. {\begin{array}{l}
 r_{ms} \equiv M\chi _{ms}^2 , \\
 \chi _{ms} = \left\{ {3 + A_2 \pm \left[ {\left( {3 - A_1 } \right)\left(
{3 + A_1 + 2A_2 } \right)} \right]^{1 / 2}} \right\}^{1 / 2},\mbox{ } \\
 A_1 = 1 + \left( {1 - a_ * ^2 } \right)^{1 / 3}\left[ {\left( {1 + a_ * }
\right)^{1 / 3} + \left( {1 - a_ * } \right)^{1 / 3}} \right], \\
 A_2 = \left( {3a_ * ^2 + A_1^2 } \right)^{1 / 2} \\
 \end{array}} \right\}
\end{equation}

\quad\quad It is found, from the criterion (\ref{eq1}), that the
screw instability will occur, provided that the BH spin $a_ * $
and the power-law index $n$ are great enough. For the given values
of $a_
* $ and $n$ we can determine the disk region for the screw
instability by using the criterion (\ref{eq1}) as follows,


\begin{equation}
\label{eq4} \xi _S < \xi < \infty ,
\end{equation}

\noindent where $\xi _S \equiv {r_{_{S}} } \mathord{\left/
{\vphantom {{r_{_{S}} } {r_{ms} }}} \right.
\kern-\nulldelimiterspace} {r_{ms} }$ is the minimum radial
coordinate for the screw instability. We think that the screw
instability prevents the closed field lines to reach the region
indicated by inequality (\ref{eq4}). It seems reasonable that the
configuration of the magnetic field in this region might consist
of the open field lines connecting the disk with the astrophysical
load, and this region is referred to as the DL region
corresponding to the DL process. Considering the restriction of
the screw instability, we modify the configuration of the magnetic
field as shown in Figure 1, and the three mechanisms of extracting
energy magnetically from the BH accretion disk are included:
(\ref{eq1}) the BZ process involved the open field lines
connecting the BH with the astrophysical load, (\ref{eq2}) the MC
process involved the closed field lines connecting the BH with the
surrounding disk, and (\ref{eq3}) the DL process involved the open
field lines connecting the disk with the astrophysical load.


\begin{figure*}
\vspace{0.5cm}
\begin{center}
\includegraphics[width=10cm]{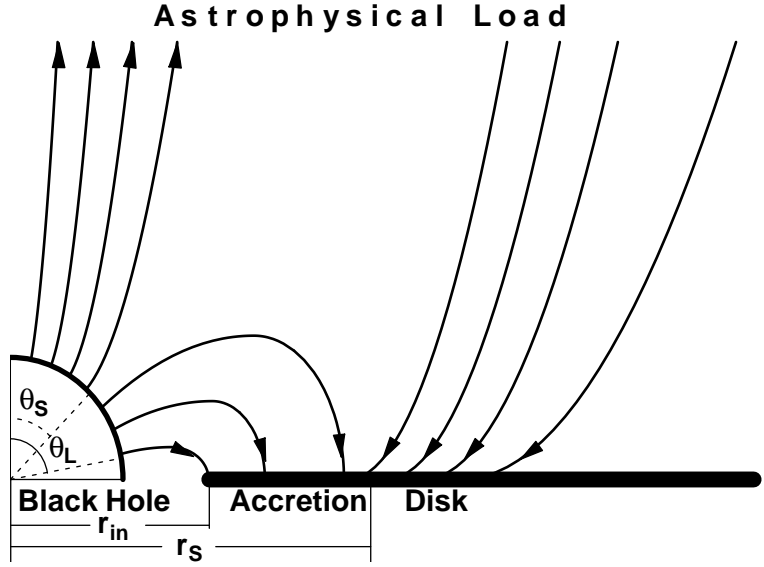}
\caption{ Configuration of the magnetic field restricted by the
screw instability}\label{fig1}
\end{center}
\end{figure*}

\quad\quad Now we give a brief description for the configuration
in Figure 1. In W03b we discussed the angular boundary between the
open and closed field lines on the BH horizon. It is shown that
the boundary angle $\theta _M $ exists in CEBZMC, if the
parameters $a_ * $ and $n$ are great enough. In W04 we argued that
the state of CEBZMC always accompanies the screw instability, and
the minimum radial coordinate $\xi _S \equiv {r_{_{S}} }
\mathord{\left/ {\vphantom {{r_{_{S}} } {r_{ms} }}} \right.
\kern-\nulldelimiterspace} {r_{ms} }$ for the screw instability
can be determined by the criterion (\ref{eq1}). Considering the
restriction of the screw instability to the closed field lines and
the action of the magnetic pressure on the horizon, we think that
the boundary angle will be extended from $\theta _M $ to $\theta
_S $. The angles $\theta _M $ and $\theta _S $ are related
respectively to infinitive and the minimum radial coordinate $\xi
_S $ by the mapping relation (W04):


\begin{equation}
\label{eq5} \cos \theta - \cos \theta _L = \int_1^\xi
{\mbox{G}\left( {a_ * ;\xi ,n} \right)d\xi } ,
\end{equation}

\noindent where


\begin{equation}
\label{eq6} \mbox{G}\left( {a_ * ;\xi ,n} \right) = \frac{\xi ^{1
- n}\chi _{ms}^2 \sqrt {1 + a_ * ^2 \chi _{ms}^{ - 4} \xi ^{ - 2}
+ 2a_ * ^2 \chi _{ms}^{ - 6} \xi ^{ - 3}} }{2\sqrt {\left( {1 + a_
* ^2 \chi _{ms}^{ - 4} + 2a_ * ^2 \chi _{ms}^{ - 6} }
\right)\left( {1 - 2\chi _{ms}^{ - 2} \xi ^{ - 1} + a_ * ^2 \chi
_{ms}^{ - 4} \xi ^{ - 2}} \right)} }.
\end{equation}

Therefore the angular region of the open field lines for the BZ
process is given by


\begin{equation}
\label{eq7} 0 < \theta < \theta _S ,
\end{equation}

\noindent and the angular region of the closed field lines for the
MC process is given by


\begin{equation}
\label{eq8} \theta _S < \theta < \theta _L ,
\end{equation}

\noindent where $\theta _L $ is the lower boundary angle of the
closed field lines. Throughout this paper $\theta _L = 0.45\pi $
is assumed in calculations. Accordingly the value range of the
radial coordinate for the MC process is given by


\begin{equation}
\label{eq9} 1 < \xi < \xi _S ,
\end{equation}

\noindent where $\xi = 1$ and $\xi _S $ correspond to $\theta _L $
and $\theta _S $ by the mapping relation (\ref{eq5}),
respectively.

\quad\quad Following Blandford (1976), we assume that the poloidal
magnetic field varies with the parameter $\xi $ on the disk as a
power law (W03a, W03b),


\begin{equation}
\label{eq10} B_D^p = B_H^p \left[ {{r_{_{H}} } \mathord{\left/
{\vphantom {{r_{_{H}} } {\varpi _{_{D}} \left( {r_{ms} }
\right)}}} \right. \kern-\nulldelimiterspace} {\varpi _{_{D}}
\left( {r_{ms} } \right)}} \right]\xi ^{ - n}, \quad 1 < \xi < \xi
_S ,
\end{equation}

\noindent where $B_H^p $ is the poloidal component of the magnetic
field on the horizon, and $n$ is the power-law index of $B_D^p $
varying with $\xi $. The quantities $r_H $ and $\varpi _{_{D}}
\left( {r_{ms} } \right)$ are the radius of the horizon and that
of the inner edge of the disk, respectively.

\quad\quad According to the above discussion the value range of
the radial coordinate for the DL process is given by equation
(\ref{eq4}), and the corresponding poloidal magnetic field varies
with the parameter $\xi $ as follows,


\begin{equation}
\label{eq11} B_D^p = B_H^p \left( {{r_{_{H}} } \mathord{\left/
{\vphantom {{r_{_{H}} } {r_{_{S}} }}} \right.
\kern-\nulldelimiterspace} {r_{_{S}} }} \right)(\xi / \xi _S )^{ -
n}, \quad \xi _S < \xi < \infty .
\end{equation}

Equation (\ref{eq11}) is similar to equation (\ref{eq10}), where
$\varpi _{_{D}} \left( {r_{ms} } \right)$ and $\xi _{in} = 1$ are
replaced by $r_{_{S}} $ and $\xi _S $, respectively.


\section{POWER OF MAGNETIC EXTRACTION FROM BH ACCRETION DISK}

\quad\quad It is tempting for us to discuss the power extracted
from BH accretion disk based on the above magnetic field
configuration. In W02 we derived the expressions for the BZ and MC
powers by using an equivalent circuit as shown in Figure 2
(henceforth circuit I), where segments $P_1 S_1 $ and $Q_1 R_1 $
represent the two adjacent magnetic surfaces, and segments $P_1
Q_1 $ and $R_1 S_1 $ represent the BH horizon and the loads
(either the astrophysical load or the disk load) sandwiched by the
two magnetic surfaces, respectively.


\begin{figure}
\vspace{0.5cm}
\begin{center}
\includegraphics[width=10cm]{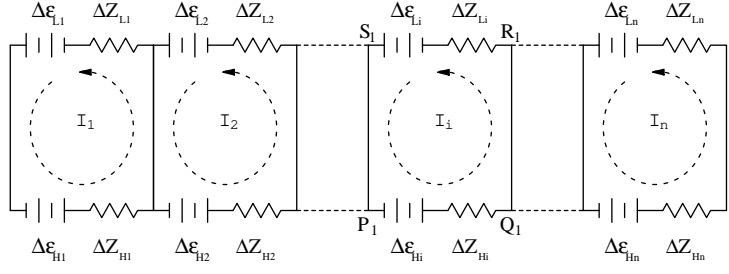}
\caption{  Equivalent circuit for the BZ and MC processes (circuit
I)}\label{fig2}
\end{center}
\end{figure}

\quad\quad By using circuit I we derive the expressions for the BZ
and MC powers and those for the BZ and MC torques corresponding to
the modified configuration of the magnetic field as follows,


\begin{equation}
\label{eq12} {P_{BZ} } \mathord{\left/ {\vphantom {{P_{BZ} } {P_0
}}} \right. \kern-\nulldelimiterspace} {P_0 } = 2a_ * ^2
\int_0^\theta {\frac{k\left( {1 - k} \right)\sin ^3\theta d\theta
}{2 - \left( {1 - q} \right)\sin ^2\theta }} , \quad 0 < \theta <
\theta _S
\end{equation}


\begin{equation}
\label{eq13} {T_{BZ} } \mathord{\left/ {\vphantom {{T_{BZ} } {T_0
}}} \right. \kern-\nulldelimiterspace} {T_0 } = 4a_ * \left( {1 +
q} \right)\int_0^{\theta _S } {\frac{\left( {1 - k} \right)\sin
^3\theta d\theta }{2 - \left( {1 - q} \right)\sin ^2\theta }} ,
\quad 0 < \theta < \theta _S
\end{equation}


\begin{equation}
\label{eq14} {P_{MC} } \mathord{\left/ {\vphantom {{P_{MC} } {P_0
}}} \right. \kern-\nulldelimiterspace} {P_0 } = 2a_ * ^2
\int_{\theta _S }^\theta {\frac{\beta \left( {1 - \beta }
\right)\sin ^3\theta d\theta }{2 - \left( {1 - q} \right)\sin
^2\theta }} , \quad \theta _S < \theta < \theta _L
\end{equation}


\begin{equation}
\label{eq15} {T_{MC} } \mathord{\left/ {\vphantom {{T_{MC} } {T_0
}}} \right. \kern-\nulldelimiterspace} {T_0 } = 4a_ * \left( {1 +
q} \right)\int_{\theta _S }^\theta {\frac{\left( {1 - \beta }
\right)\sin ^3\theta d\theta }{2 - \left( {1 - q} \right)\sin
^2\theta }} , \quad \theta _S < \theta < \theta _L
\end{equation}

\noindent where $\Delta P_{BZ} $, $\Delta T_{BZ} $, $\Delta P_{MC}
$ and $\Delta T_{MC} $ are related by


\begin{equation}
\label{eq16} \Delta P_{BZ} = \Omega _F \Delta T_{BZ} , \quad
\Delta P_{MC} = \Omega _D \Delta T_{MC} .
\end{equation}

\quad\quad In equations (\ref{eq12})---(\ref{eq15}) the parameters
$k$ and $\beta $ are respectively the ratios of the angular
velocities of the open and closed field lines to the angular
velocity of the horizon and read


\begin{equation}
\label{eq17} k \equiv {\Omega _F } \mathord{\left/ {\vphantom
{{\Omega _F } {\Omega _H }}} \right. \kern-\nulldelimiterspace}
{\Omega _H }, \quad \beta \equiv {\Omega _D } \mathord{\left/
{\vphantom {{\Omega _D } {\Omega _H }}} \right.
\kern-\nulldelimiterspace} {\Omega _H } = \frac{2\left( {1 + q}
\right)}{a_ * }\left[ {\left( {\sqrt \xi \chi _{ms} } \right)^3 +
a_ * } \right]^{ - 1}.
\end{equation}

\noindent where $\Omega _F $ is the angular velocity of the
magnetic field line, $\Omega _H $ and $\Omega _D $ are
respectively the angular velocities of the horizon and the disk
and read


\begin{equation}
\label{eq18} \Omega _H = \frac{a_ * }{2r_H }, \quad \Omega _D =
\frac{1}{M(\xi ^{3 \mathord{\left/ {\vphantom {3 2}} \right.
\kern-\nulldelimiterspace} 2}\chi _{ms}^3 + a_ * )}.
\end{equation}

\quad\quad Usually $k = 0.5$ is taken for the optimal BZ power
(MT82), while $\beta $ depends on the BH spin and the place where
the field line penetrates on the disk. In equations
(\ref{eq12})---(\ref{eq15}) we use the parameter $q \equiv \sqrt
{1 - a_
* ^2 } $, and the parameters $P_0 $ and $T_0 $ are defined as


\begin{equation}
\label{eq19} \left\{ {\begin{array}{l}
 P_0 \equiv \left( {B_H^p } \right)^2M^2 \approx B_4^2 \left( {M
\mathord{\left/ {\vphantom {M {M_ \odot }}} \right.
\kern-\nulldelimiterspace} {M_ \odot }} \right)^2\times 6.59\times
10^{28}erg \cdot s^{ - 1}, \\
 T_0 \equiv \left( {B_H^p } \right)^2M^3 \approx B_4^2 \left( {M
\mathord{\left/ {\vphantom {M {M_ \odot }}} \right.
\kern-\nulldelimiterspace} {M_ \odot }} \right)^3\times 3.26\times
10^{23}g
\cdot cm^2 \cdot s^{ - 2}, \\
 \end{array}} \right.
\end{equation}

\noindent where $B_4 $ is the strength of the poloidal magnetic
field on the horizon in units of $10^4\mbox{ }gauss$.

\quad\quad Considering that the poloidal magnetic field will exert
a braking torque on the current in the DL region, we can calculate
the DL power and torque by using the equivalent circuit as shown
in Figure 3. Henceforth this equivalent circuit is referred to as
circuit II, where segments $P_2 S_2 $ and $Q_2 R_2 $ represent the
two adjacent magnetic surfaces consisting of the open field lines
in the DL region, and segments $P_2 Q_2 $ and $R_2 S_2 $ represent
the disk surface in the DL region and the load sandwiched by the
two adjacent magnetic surfaces, respectively. The
quantities\textbf{ }$\Delta Z_A $ and $\Delta \varepsilon _{_{D}}
$ (the subscript ``$i$'' is omitted) are the resistance of the
load and EMF due to the rotation of the disk, respectively. The
disk load is neglected in calculations by considering the perfect
conductivity of plasma.


\begin{figure}
\vspace{0.5cm}
\begin{center}
\includegraphics[width=10cm]{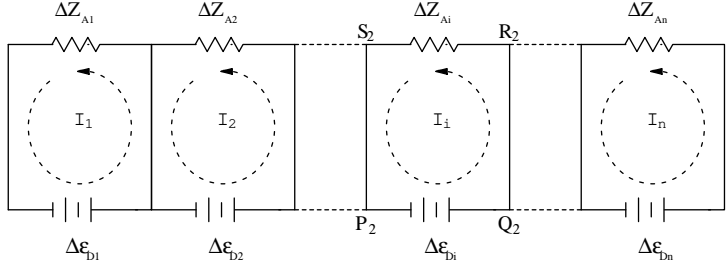}
\caption{ Equivalent circuit for the DL process (circuit
II)}\label{fig3}
\end{center}
\end{figure}

\quad\quad The following equations are used to derive the DL power
in circuit II, which are similar to those given in deriving the BZ
power and MC power in W02.


\begin{equation}
\label{eq20} \Delta P_{DL} = I_{DL}^2 \Delta Z_A , \quad I_{DL} =
{\Delta \varepsilon _{_{D}} } \mathord{\left/ {\vphantom {{\Delta
\varepsilon _{_{D}} } {\Delta Z_A }}} \right.
\kern-\nulldelimiterspace} {\Delta Z_A }, \quad \Delta \varepsilon
_{_{D}} = - \left( {{\Delta \Psi _D } \mathord{\left/ {\vphantom
{{\Delta \Psi _D } {2\pi }}} \right. \kern-\nulldelimiterspace}
{2\pi }} \right)\Omega _D ,
\end{equation}

\noindent where $I_{DL} $ is the current in each loop of circuit
II. The minus sign in $\Delta \varepsilon _D $ arises from the
direction of the magnetic flux between the two adjacent magnetic
surfaces, which is given by


\begin{equation}
\label{eq21} \Delta \Psi _D = B_D^P 2\pi \left( {{\varpi \rho }
\mathord{\left/ {\vphantom {{\varpi \rho } {\sqrt \Delta }}}
\right. \kern-\nulldelimiterspace} {\sqrt \Delta }}
\right)_{\theta = \pi \mathord{\left/ {\vphantom {\pi 2}} \right.
\kern-\nulldelimiterspace} 2} dr.
\end{equation}

The concerned Kerr metric coefficients are given by (Thorne, Price
{\&} MacDonald 1986)


\begin{equation}
\label{eq22} \left\{ {\begin{array}{l}
 \varpi = \left( {\Sigma \mathord{\left/ {\vphantom {\Sigma \rho }} \right.
\kern-\nulldelimiterspace} \rho } \right)\sin \theta , \\
 \Sigma ^2 \equiv \left( {r^2 + a^2} \right)^2 - a^2\Delta \sin ^2\theta ,
\\
 \rho ^2 \equiv r^2 + a^2\cos ^2\theta , \\
 \Delta \equiv r^2 + a^2 - 2Mr. \\
 \end{array}} \right.
\end{equation}

\quad\quad Since the astrophysical load remains unknown, we give
some simplified assumptions as follows.

\quad\quad (\ref{eq1}) The load is axisymmetric, being located
evenly in a plane\textbf{\textit{ P}} with some height above the
disk.

\quad\quad (\ref{eq2}) Each open field line intersects with the
disk and the plane \textbf{\textit{P }}at the cylindrical radii
$r$ and ${r}'$, respectively. The both radii are related by


\begin{equation}
\label{eq23} {r}' = \lambda r,
\end{equation}

\noindent where $\lambda $ is assumed to be a constant.

\quad\quad (\ref{eq3}) The surface resistivity $\sigma _{_{L}} $
of the unknown load is assumed to obey the following relation,


\begin{equation}
\label{eq24} \sigma _{_{L}} = \alpha _{_{Z}} \sigma _{_{H}} = 4\pi
\alpha _{_{Z}} ,
\end{equation}

\noindent where $\alpha _{_{Z}} $ is a parameter, and $\sigma
_{_{H}} = 4\pi = 377\mbox{ }ohm$ is the surface resistivity of the
BH horizon. Throughout this paper $\alpha _{_{Z}} = 1$ is assumed
in calculations. The load resistance $\Delta Z_A $ between the two
adjacent magnetic surfaces can be written as


\begin{equation}
\label{eq25} \Delta Z_A = \sigma _{_{L}} \frac{d{r}'}{2\pi {r}'} =
2\alpha _{_{Z}} \frac{dr}{r},
\end{equation}

where equations (\ref{eq23}) and (\ref{eq24}) are used in the last
step. Incorporating equation (\ref{eq11}) with equations
(\ref{eq20})\mbox{---}(\ref{eq25}), we have


\begin{equation}
\label{eq26} {\Delta P_{DL} } \mathord{\left/ {\vphantom {{\Delta
P_{DL} } {P_0 }}} \right. \kern-\nulldelimiterspace} {P_0 } =
f\left( {a_ * ,\xi ,n} \right)d\xi ,
\end{equation}

\noindent where the function $ f\left( {a_ * ,\xi ,n} \right)$ is
expressed by


\begin{equation}
\label{eq27}f(a_\ast ,\xi ,n) = \frac{\left( {1 + q}
\right)^2\left( {1 + a_\ast ^2 \chi _{ms}^{ - 4} \xi ^{ - 2} +
2a_\ast ^2 \chi _{ms}^{ - 6} \xi ^{ - 3}} \right)\chi _{ms}^4 \xi
_S \left( {\xi \mathord{\left/ {\vphantom {\xi {\xi _S }}} \right.
\kern-\nulldelimiterspace} {\xi _S }} \right)^{ - 2n + 3}}{2\left(
{\xi ^{3 / 2}\chi _{ms}^3 + a_\ast } \right)^2\left( {1 - 2\chi
_{ms}^{ - 2} \xi ^{ - 1} + a_\ast ^2 \chi _{ms}^{ - 4} \xi ^{ -
2}} \right)}.
\end{equation}

Similarly, the DL power is related to the DL torque by


\begin{equation}
\label{eq28} \Delta P_{DL} = \Omega _D \Delta T_{DL} ,
\end{equation}

\noindent and we have


\begin{equation}
\label{eq29} {\Delta T_{DL} } \mathord{\left/ {\vphantom {{\Delta
T_{DL} } {T_0 }}} \right. \kern-\nulldelimiterspace} {T_0 } =
g\left( {a_ * ,\xi ,n} \right)d\xi ,
\end{equation}

\noindent where the function $g\left( {a_ * ,\xi ,n} \right)$ is
expressed by


\begin{equation}
\label{eq30} g(a_\ast ,\xi ,n) = \frac{\left( {1 + q}
\right)^2\left( {1 + a_\ast ^2 \chi _{ms}^{ - 4} \xi ^{ - 2} +
2a_\ast ^2 \chi _{ms}^{ - 6} \xi ^{ - 3}} \right)\chi _{ms}^4 \xi
_S \left( {\xi \mathord{\left/ {\vphantom {\xi {\xi _S }}} \right.
\kern-\nulldelimiterspace} {\xi _S }} \right)^{ - 2n + 3}}{2\left(
{\xi ^{3 / 2}\chi _{ms}^3 + a_\ast } \right)\left( {1 - 2\chi
_{ms}^{ - 2} \xi ^{ - 1} + a_\ast ^2 \chi _{ms}^{ - 4} \xi ^{ -
2}} \right)}.
\end{equation}

Integrating equations (\ref{eq26}) and (\ref{eq29}) over the DL
region, we have the DL power and torque expressed by


\begin{equation}
\label{eq31} {P_{DL} (a_\ast ,\xi ,n)} \mathord{\left/ {\vphantom
{{P_{DL} (a_\ast ,\xi ,n)} {P_0 }}} \right.
\kern-\nulldelimiterspace} {P_0 } = \int_{\xi _S }^\xi {f(a_\ast
,{\xi }',n)} d{\xi }',
\end{equation}


\begin{equation}
\label{eq32} {T_{DL} \left( {a_ * ,\xi ,n} \right)}
\mathord{\left/ {\vphantom {{T_{DL} \left( {a_ * ,\xi ,n} \right)}
{T_0 }}} \right. \kern-\nulldelimiterspace} {T_0 } = \int_{\xi _S
}^\xi g \left( {a_ * ,{\xi }',n} \right)d{\xi }'.
\end{equation}

\quad\quad For the given values of $a_ * $ and $n$ the strength of
${P_{BZ} } \mathord{\left/ {\vphantom {{P_{BZ} } {P_0 }}} \right.
\kern-\nulldelimiterspace} {P_0 }$, ${P_{MC} } \mathord{\left/
{\vphantom {{P_{MC} } {P_0 }}} \right. \kern-\nulldelimiterspace}
{P_0 }$ and ${P_{DL} } \mathord{\left/ {\vphantom {{P_{DL} } {P_0
}}} \right. \kern-\nulldelimiterspace} {P_0 }$ are compared by
using equations (\ref{eq12}), (\ref{eq14}) and (\ref{eq31}) as
shown in Figures 4.


\begin{figure}
\vspace{0.5cm}
\begin{center}
{\includegraphics[width=6cm]{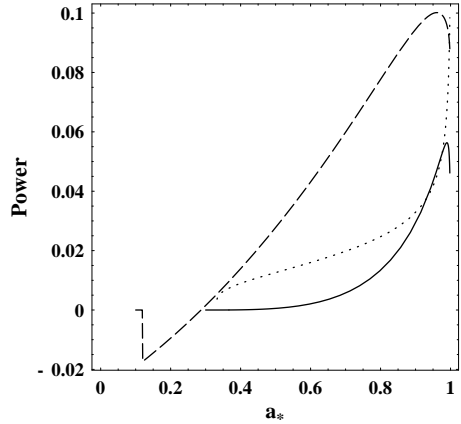}
 \centerline{\quad\quad\quad(a)}
 \includegraphics[width=6cm]{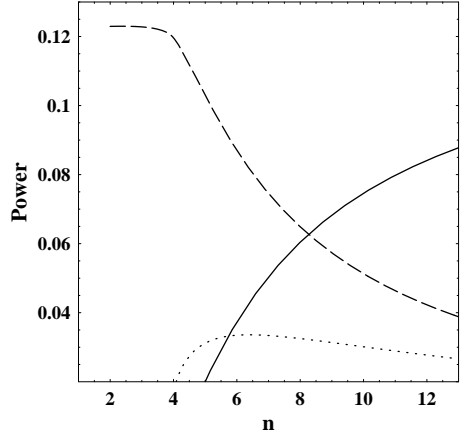}
 \centerline{\quad\quad\quad(b)}}
 \caption{ The BZ power (solid line), the MC power
(dashed line) and the DL power (dotted line) (a) versus the BH
spin $a_ * $ with $n = 5.5$, and (b) versus the power-law index
$n$ with $a_ * = 0.9$.}\label{fig4}
\end{center}
\end{figure}

From Figure 4 we obtain the following results,

\quad\quad (\ref{eq1}) For the increasing $a_ * $ with the given
$n$, both the BZ and the MC powers vary non-monotonically,
attaining their maxima as the BH spin approaches unity, while the
DL power increases monotonically.

\quad\quad (\ref{eq2}) For the increasing $n$ with the given $a_ *
$, the BZ power increases monotonically, the MC power decreases
monotonically, while the DL power varies non-monotonically,
attaining its maximum value $P_{DL} \approx 0.034$ with $n =
6.39$.

\quad\quad (\ref{eq3}) The MC power is generally greater than the
BZ power, if the power-law index $n$ is not very big. The DL power
is generally less than the BZ and MC powers, and it is comparable
with the latter two powers as the BH spin approaches unity.


\section{RADIATION FROM A MAGNETIZED ACCRETION DISK }

\subsection{Radiation from the MC region}

\quad\quad Radiation from a relativistic steady thin disk around a
Kerr BH has been discussed based on the three conservation laws of
mass, energy and angular momentum with the ``no-torque boundary
condition'' by some authors (Novikov {\&} Thorne 1973, Page {\&}
Thorne 1974). By combining the MC effects with the above
conservation laws and the ``no-torque boundary condition'', the
following equation of radiation from a relativistic quasi-steady
thin disk was derived in L02,


\begin{equation}
\label{eq33} F_{MC}^{total} = F_{DA}^I + F_{MC} ,
\end{equation}

\noindent where $F_{DA}^I $ and $F_{MC} $ are the radiation fluxes
due to disk accretion and the MC effects, respectively, and they
are expressed by


\begin{equation}
\label{eq34} F_{DA}^I = \frac{\dot {M}_D }{4\pi r}\frac{\left( {{
- \partial \Omega _D } \mathord{\left/ {\vphantom {{ - \partial
\Omega _D } {\partial r}}} \right. \kern-\nulldelimiterspace}
{\partial r}} \right)}{\left( {E^ + - \Omega _D L^ + }
\right)^2}\int_{r_{ms} }^r {\left( {E^ + - \Omega _D L^ + }
\right)\left( {{\partial L^ + } \mathord{\left/ {\vphantom
{{\partial L^ + } {\partial r}}} \right.
\kern-\nulldelimiterspace} {\partial r}} \right)dr} ,
\end{equation}


\begin{equation}
\label{eq35} F_{MC} = \frac{\left( {{ - \partial \Omega _D }
\mathord{\left/ {\vphantom {{ - \partial \Omega _D } {\partial
r}}} \right. \kern-\nulldelimiterspace} {\partial r}}
\right)}{r\left( {E^ + - \Omega _D L^ + } \right)^2}\int_{r_{ms}
}^r {\left( {E^ + - \Omega _D L^ + } \right)rH_{MC} dr} .
\end{equation}

In equation (\ref{eq35}) $H_{MC} $ is the flux of angular momentum
transferred between the BH and the disc by the magnetic field, and
$E^ + $ and $L^ + $ are the specific energy and angular momentum
of a particle in the disc, respectively, and they read (Novikov
{\&} Thorne 1973)


\begin{equation}
\label{eq36} E^ + = {\left( {1 - 2\chi ^{ - 2} + a_ * \chi ^{ -
3}} \right)} \mathord{\left/ {\vphantom {{\left( {1 - 2\chi ^{ -
2} + a_ * \chi ^{ - 3}} \right)} {\left( {1 - 3\chi ^{ - 2} + 2a_
* \chi ^{ - 3}} \right)^{1 / 2}}}} \right.
\kern-\nulldelimiterspace} {\left( {1 - 3\chi ^{ - 2} + 2a_ * \chi
^{ - 3}} \right)^{1 / 2}},
\end{equation}


\begin{equation}
\label{eq37} L^ + = M\chi {\left( {1 - 2a_ * \chi ^{ - 3} + a_ *
^2 \chi ^{ - 4}} \right)} \mathord{\left/ {\vphantom {{\left( {1 -
2a_ * \chi ^{ - 3} + a_ * ^2 \chi ^{ - 4}} \right)} {\left( {1 -
3\chi ^{ - 2} + 2a_ * \chi ^{ - 3}} \right)^{1 / 2}}}} \right.
\kern-\nulldelimiterspace} {\left( {1 - 3\chi ^{ - 2} + 2a_ * \chi
^{ - 3}} \right)^{1 / 2}}.
\end{equation}

From equations (\ref{eq36}) and (\ref{eq37}) we have $E^ + =
E_{ms} $ and $L^ + = L_{ms} $ for $\xi = 1$ with $\chi = \chi
_{ms} $. The expression for $H_{MC} $ can be worked out by using
the following relation,


\begin{equation}
\label{eq38} {\partial T_{MC} } \mathord{\left/ {\vphantom
{{\partial T_{MC} } {\partial r}}} \right.
\kern-\nulldelimiterspace} {\partial r} = \frac{\left( {{\partial
T_{MC} } \mathord{\left/ {\vphantom {{\partial T_{MC} } {\partial
\theta }}} \right. \kern-\nulldelimiterspace} {\partial \theta }}
\right)\left( {{d\theta } \mathord{\left/ {\vphantom {{d\theta }
{d\xi }}} \right. \kern-\nulldelimiterspace} {d\xi }} \right)}{\xi
M\chi _{ms}^2 } = 4\pi rH_{MC} ,
\end{equation}

\noindent where ${\partial T_{MC} } \mathord{\left/ {\vphantom
{{\partial T_{MC} } {\partial \theta }}} \right.
\kern-\nulldelimiterspace} {\partial \theta }$ and ${d\theta }
\mathord{\left/ {\vphantom {{d\theta } {d\xi }}} \right.
\kern-\nulldelimiterspace} {d\xi }$ can be calculated using
equations (\ref{eq15}) and (\ref{eq5}). In W03a we expressed
$H_{MC} $ as follows,


\begin{equation}
\label{eq39} {H_{MC} } \mathord{\left/ {\vphantom {{H_{MC} } {H_0
}}} \right. \kern-\nulldelimiterspace} {H_0 } = A\left( {a_ * ,\xi
} \right)\xi ^{ - n}, \quad\quad 1 < \xi  < \xi _{S} ,
\end{equation}

\noindent where


\begin{equation}
\label{eq40} H_0 = \left( {B_H^p } \right)^2M = 1.48\times
10^{13}\times B_4^2 \left( {M \mathord{\left/ {\vphantom {M {M_
\odot }}} \right. \kern-\nulldelimiterspace} {M_ \odot }}
\right)\mbox{ }g \cdot s^{ - 2},
\end{equation}


\begin{equation}
\label{eq41} \left\{ {\begin{array}{l}
 A\left( {a_ * ,\xi } \right) = \frac{a_ * \left( {1 - \beta } \right)\left(
{1 + q} \right)}{2\pi \chi _{ms}^2 \left[ {2\csc ^2\theta - \left(
{1 - q}
\right)} \right]}F_H \left( {a_ * ,\xi } \right), \\\\
 F_H \left( {a_ * ,\xi } \right) = \frac{\sqrt {1 + a_ * ^2 \chi _{ms}^{ -
4} \xi ^{ - 2} + 2a_ * ^2 \chi _{ms}^{ - 6} \xi ^{ - 3}} }{\sqrt
{\left( {1 + a_ * ^2 \chi _{ms}^{ - 4} + 2a_ * ^2 \chi _{ms}^{ -
6} } \right)\left( {1 - 2\chi _{ms}^{ - 2} \xi ^{ - 1} + a_ * ^2
\chi _{ms}^{ - 4} \xi ^{ - 2}}
\right)} }. \\
 \end{array}} \right.
\end{equation}

\quad\quad Since the magnetic field on the horizon is brought and
held by its surrounding magnetized disk, there must exist some
relations between the magnetic field and the accretion rate. As a
matter of fact these relations might be rather complicated, and
would be very different in different situations. One of them is
given by considering the balance between the pressure of the
magnetic field on the horizon and the ram pressure of the
innermost parts of an accretion flow (Moderski, Sikora {\&} Lasota
1997), i.e.,


\begin{equation}
\label{eq42} {\left( {B_H^p } \right)^2} \mathord{\left/
{\vphantom {{\left( {B_H^p } \right)^2} {\left( {8\pi } \right)}}}
\right. \kern-\nulldelimiterspace} {\left( {8\pi } \right)} =
P_{ram} \sim \rho c^2\sim {\dot {M}_D } \mathord{\left/ {\vphantom
{{\dot {M}_D } {\left( {4\pi r_H^2 } \right)}}} \right.
\kern-\nulldelimiterspace} {\left( {4\pi r_H^2 } \right)},
\end{equation}

From equation (\ref{eq42}) we define $F_0 $ as


\begin{equation}
\label{eq43} F_0 \equiv \left( {B_H^p } \right)^2 = \frac{2\dot
{M}_D }{M^2\left( {1 + q} \right)^2}.
\end{equation}

\quad\quad By using equations (\ref{eq34}) and (\ref{eq35}) we
have the radiation fluxes, ${F_{DA}^I } \mathord{\left/ {\vphantom
{{F_{DA}^I } {F_0 }}} \right. \kern-\nulldelimiterspace} {F_0 }$
and ${F_{MC} } \mathord{\left/ {\vphantom {{F_{MC} } {F_0 }}}
\right. \kern-\nulldelimiterspace} {F_0 }$, versus the radial
parameter $\xi $ for the given values of the power-law index and
the BH spin as shown in Figure 5.

\begin{figure}
\vspace{0.5cm}
\begin{center}
{\includegraphics[width=6cm]{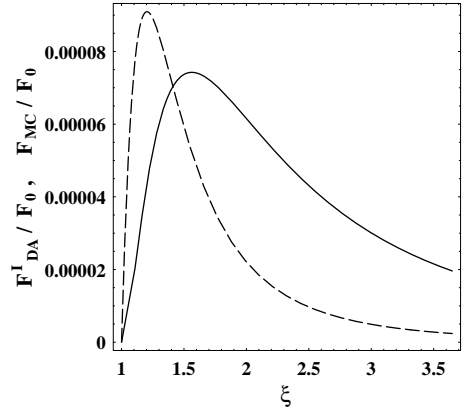}
 \centerline{\quad\quad\quad\quad(a)}
 \includegraphics[width=6cm]{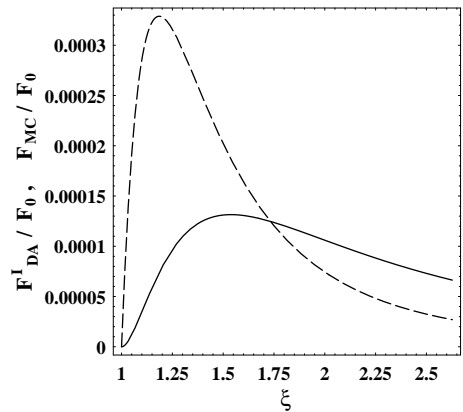}
 \centerline{\quad\quad\quad\quad(b)}
 \includegraphics[width=6cm]{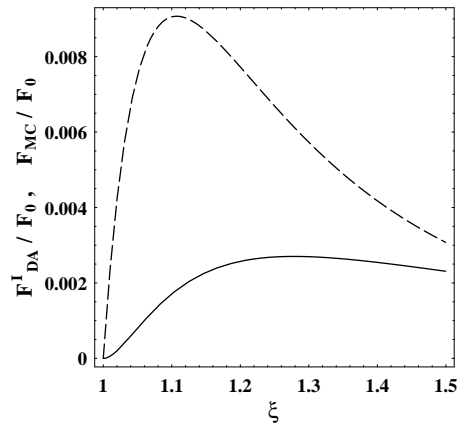}
 \centerline{\quad\quad\quad\quad(c)}}
 \caption{The radiation flux from the MC region, $F_{DA}^I  / F_0 $ (solid
line) and $F_{MC}  / F_0 $ (dashed line) versus $\xi$ for $1 < \xi
< \xi _S$ with $n = 5.5$ and different values of the BH spin: (a)
$a_ * = 0.5$, (b) $a_ * = 0.7$ and (c) $a_ * = 0.998$.
}\label{fig5}
\end{center}
\end{figure}

\quad\quad Expecting Figure 5 we find that the peak of ${F_{MC} }
\mathord{\left/ {\vphantom {{F_{MC} } {F_0 }}} \right.
\kern-\nulldelimiterspace} {F_0 }$ is not only closer to the inner
edge of the disk, but also is greater than that of ${F_{DA}^I }
\mathord{\left/ {\vphantom {{F_{DA}^I } {F_0 }}} \right.
\kern-\nulldelimiterspace} {F_0 }$. Furthermore, ${F_{MC} }
\mathord{\left/ {\vphantom {{F_{MC} } {F_0 }}} \right.
\kern-\nulldelimiterspace} {F_0 }$ varies more steeply with $\xi $
than ${F_{DA}^I } \mathord{\left/ {\vphantom {{F_{DA}^I } {F_0 }}}
\right. \kern-\nulldelimiterspace} {F_0 }$ does. Thus the
radiation flux due to the MC mechanism can result in a very steep
emissivity in the inner region of the disk, which is consistent
with the recent \textit{XMM-Newton} observation of the nearby
bright Seyfert 1 galaxy MCG-6-30-15 (Wilms et al. 2001; Li 2002b;
W03a).

\subsection{Radiation from the DL region}

\quad\quad We can show that the radiation flux from the DL region
also consists of two terms:


\begin{equation}
\label{eq44} F_{DL}^{Total} = F_{DA}^{II} + F_{DL} ,
\end{equation}

\noindent where $F_{DL} $ is the electromagnetic flux in the DL
process, and $F_{DA}^{II} $ is the radiation flux due to disk
accretion in the DL region. The flux $F_{DL} $ can be worked out
by


\begin{equation}
\label{eq45} F_{DL} = \frac{\Delta P_{DL} }{2\pi r\Delta r}.
\end{equation}

Substituting equation (\ref{eq26}) into equation (\ref{eq45}), we
have


\begin{equation}
\label{eq46} {F_{DL} } \mathord{\left/ {\vphantom {{F_{DL} } {F_0
}}} \right. \kern-\nulldelimiterspace} {F_0 } = \frac{f\left( {a_
* ,\xi ,n} \right)}{2\pi \xi \chi _{ms}^4 }
\end{equation}

\quad\quad The difference between the radiation from the DL region
and that from the MC region lies in two aspects.

\quad\quad (\ref{eq1}) The outgoing flux $F_{MC} $ can be obtained
after resolving the equations of the conservation laws with
$P_{MC} $ and $T_{MC} $ incoming the disk, while the outgoing flux
$F_{DL} $ is given before resolving the equations from the
conservation laws;

\quad\quad (\ref{eq2}) ``No-torque boundary condition'' can be
used at $r_{ms} $ for the solution in the MC region, while this
boundary condition is not valid at $r_{_{S}} $ for the solution in
the DL region.

\quad\quad Thus we find the flux $F_{DA}^{II} $ by applying the
conservation laws of energy and angular momentum to the DL region,
i.e.,


\begin{equation}
\label{eq47} \frac{\partial }{\partial r}\left( {\dot {M}_D E^ + -
2\pi rW_\varphi ^r \Omega _D } \right) = 4\pi r\left( {F_{DA}^{II}
E^ + + H_{DL} \Omega _D } \right),
\end{equation}


\begin{equation}
\label{eq48} \frac{\partial }{\partial r}\left( {\dot {M}_D L^ + -
2\pi rW_\varphi ^r } \right) = 4\pi r\left( {F_{DA}^{II} L^ + +
H_{DL} } \right),
\end{equation}

\noindent where accretion rate $\dot {M}_D $ is regarded as a
constant for a quasi-steady disk, and $W_\varphi ^r $ is the
internal viscous torque per unit circumference. Incorporating
equations (\ref{eq47}) and (\ref{eq48}), we have


\begin{equation}
\label{eq49} W_\varphi ^r = \frac{2\left( {E^ + - \Omega _D L^ + }
\right)}{\left( {{ - d\Omega _D } \mathord{\left/ {\vphantom {{ -
d\Omega _D } {dr}}} \right. \kern-\nulldelimiterspace} {dr}}
\right)}F_{DA}^{II} .
\end{equation}

The quantity $H_{DL} $ is the flux of angular momentum, which is
related to the electromagnetic torque by


\begin{equation}
\label{eq50} {\partial T_{DL} } \mathord{\left/ {\vphantom
{{\partial T_{DL} } {\partial r}}} \right.
\kern-\nulldelimiterspace} {\partial r} = \left( {{\partial T_{DL}
} \mathord{\left/ {\vphantom {{\partial T_{DL} } {\partial \xi }}}
\right. \kern-\nulldelimiterspace} {\partial \xi }} \right)\left(
{{\partial \xi } \mathord{\left/ {\vphantom {{\partial \xi }
{\partial r}}} \right. \kern-\nulldelimiterspace} {\partial r}}
\right) = 4\pi rH_{DL} .
\end{equation}

Substituting equation (\ref{eq32}) into equation (\ref{eq50}), we
have


\begin{equation}
\label{eq51} {H_{DL} } \mathord{\left/ {\vphantom {{H_{DL} } {H_0
}}} \right. \kern-\nulldelimiterspace} {H_0 } = {g(a_\ast ,\xi
,n)} \mathord{\left/ {\vphantom {{g(a_\ast ,\xi ,n)} {\left( {4\pi
\xi \chi _{ms}^4 } \right)}}} \right. \kern-\nulldelimiterspace}
{\left( {4\pi \xi \chi _{ms}^4 } \right)}.
\end{equation}

Substituting equation (\ref{eq49}) into equation (\ref{eq48}) and
integrating equation (\ref{eq48}) over the DL region, we have the
expression for $F_{DA}^{II} $ as follows,


\begin{equation}
\label{eq52} F_{DA}^{II} = F_A + F_B + F_C ,
\end{equation}

\noindent where


\begin{equation}
\label{eq53} F_A = \frac{\dot {M}_D }{4\pi r}\frac{\left( {{ -
\partial \Omega _D } \mathord{\left/ {\vphantom {{ - \partial
\Omega _D } {\partial r}}} \right. \kern-\nulldelimiterspace}
{\partial r}} \right)}{\left( {E^ + - \Omega _D L^ + }
\right)^2}\int_{r_{_{S}} }^r {\left( {E^ + - \Omega _D L^ + }
\right)\left( {{\partial L^ + } \mathord{\left/ {\vphantom
{{\partial L^ + } {\partial r}}} \right.
\kern-\nulldelimiterspace} {\partial r}} \right)dr} ,
\end{equation}


\begin{equation}
\label{eq54} F_B = - \frac{\left( {{ - \partial \Omega _D }
\mathord{\left/ {\vphantom {{ - \partial \Omega _D } {\partial
r}}} \right. \kern-\nulldelimiterspace} {\partial r}}
\right)}{r\left( {E^ + - \Omega _D L^ + } \right)^2}\int_{r_{_{S}}
}^r {\left( {E^ + - \Omega _D L^ + } \right)rH_{DL} dr} ,
\end{equation}

\begin{figure}
\vspace{0.6cm}
\begin{center}
{\includegraphics[width=6.0cm]{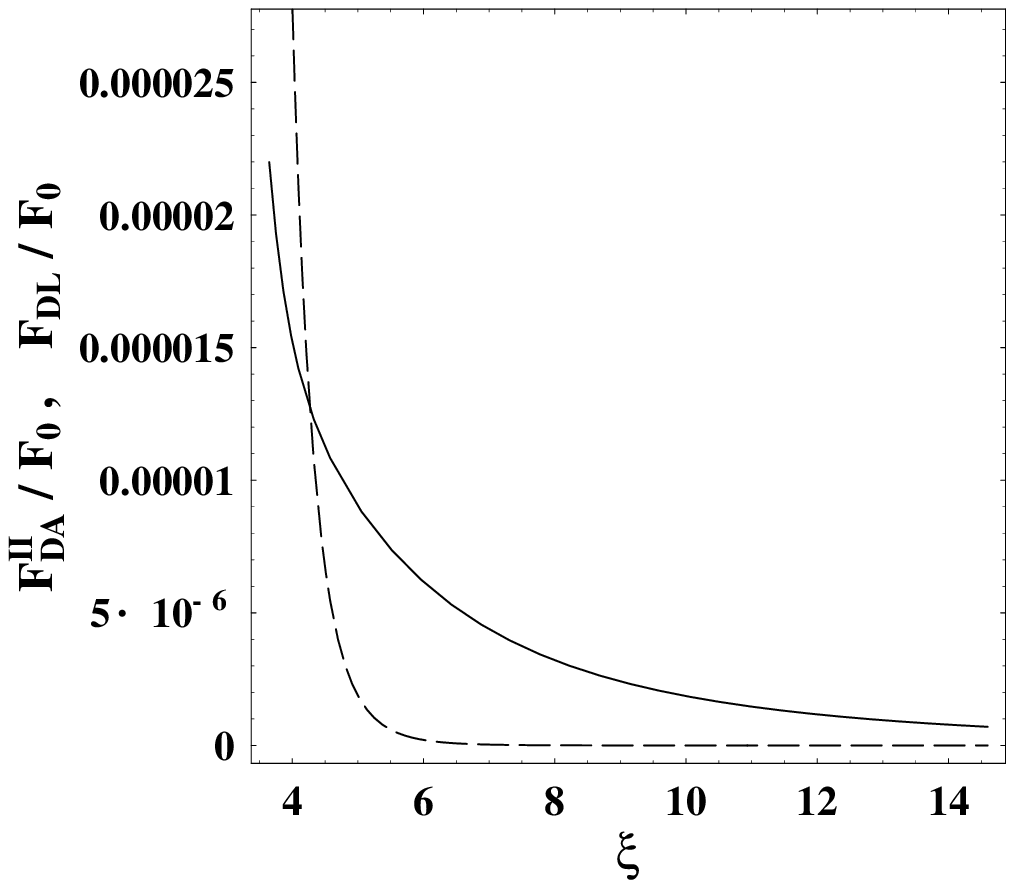}
 \centerline{\quad\quad(a)}
 \includegraphics[width=6.0cm]{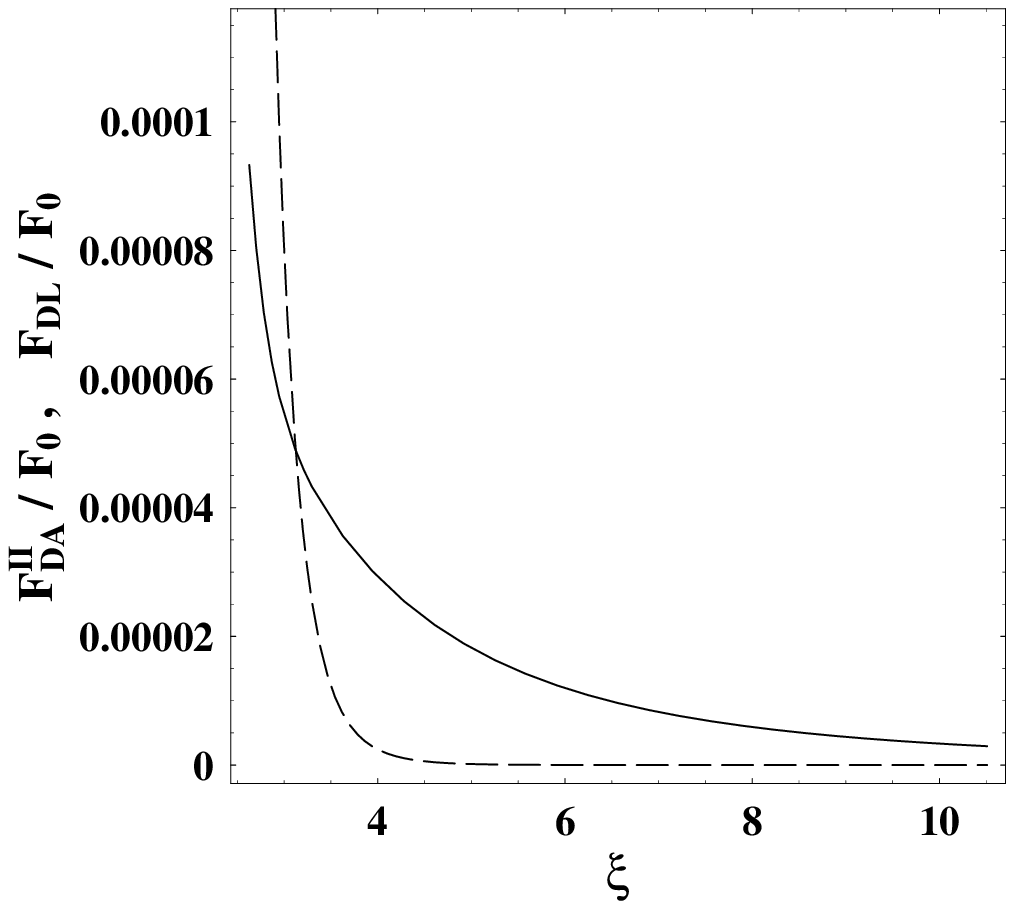}
 \centerline{\quad\quad(b)}
\includegraphics[width=6.0cm]{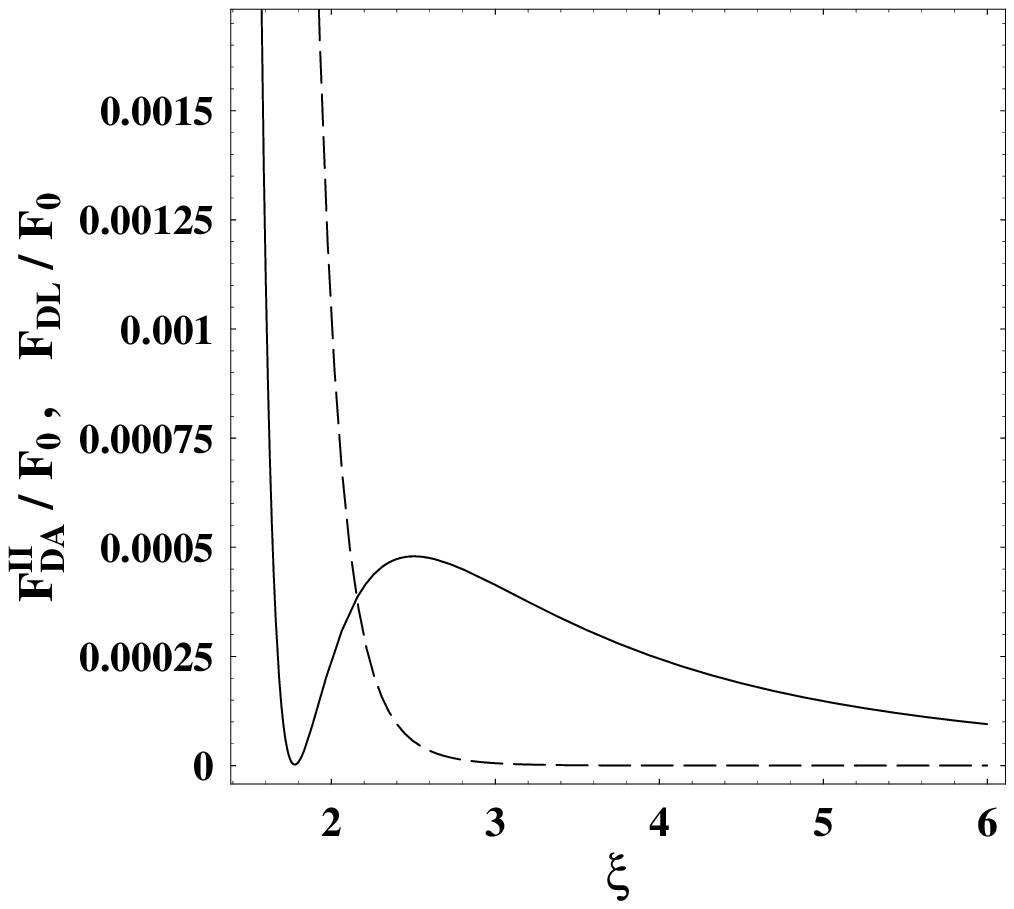}
 \centerline{\quad\quad(c)}}
\caption{ The radiation flux from the DL region, $F_{DA}^{II} /
F_0 $ (solid line) and $F_{DL} / F_0 $ (dashed line) versus $\xi$
for $\xi _S < \xi < 4 \xi _S $ with $n = 5.5$ and different values
of the BH spin: (a) $a_ * = 0.5$, (b) $a_ * = 0.7$ and (c) $a_ * =
0.998$.}\label{fig6}
\end{center}
\end{figure}


\begin{equation}
\label{eq55} F_C = \frac{- \partial \Omega_D / \partial r}{r (E^+
- \Omega_D L^+)^2}[\frac{(E^+ - \Omega_D L^+)^2 r
F_{MC}^{Total}}{- \partial \Omega_D / \partial r}]_{r = r_S^-}
\end{equation}

Expecting equations (\ref{eq52})---(\ref{eq55}), we find the
following characteristics of $F_{DA}^{II}$:

\quad\quad (\ref{eq1})The flux $F_A $ has the same form as
$F_{DA}^I $ except the integral region;

\quad\quad (\ref{eq2})The flux $F_B $ has a similar form to
$F_{MC} $ except the minus sign in equation (\ref{eq54}), which
implies the contribution of the DL torque on $F_{DA}^{II} $ is
negative;

\quad\quad (\ref{eq3})The flux $F_C $ represents the effect of the
boundary condition at $r = r_S^ - $, where ``no-torque boundary
condition'' is not valid.

\quad\quad By using equations (\ref{eq52}) and (\ref{eq46}) we
have the radiation fluxes, $F_{DA}^{II} / F_0 $ and $F_{DL} / F_0
$, versus the radial parameter $\xi $ for the given values of the
power-law index and the BH spin as shown in Figures 6.

Inspecting Figure 6, we have the following results.

\quad\quad (\ref{eq1}) The radiation flux ${F_{DL} }
\mathord{\left/ {\vphantom {{F_{DL} } {F_0 }}} \right.
\kern-\nulldelimiterspace} {F_0 }$ decreases monotonously and very
steeply with the increasing $\xi $. This result arises from two
aspects, i.e., both the angular velocity decreases and the area of
the ring from $r$ to $r + dr$ increases with the increasing $\xi
$.

\quad\quad (\ref{eq2}) The radiation flux ${F_{DA}^{II} }
\mathord{\left/ {\vphantom {{F_{DA}^{II} } {F_0 }}} \right.
\kern-\nulldelimiterspace} {F_0 }$ generally decreases
monotonously with the increasing $\xi $, while it varies
non-monotonously with $\xi $ as the BH spin approaches unity as
shown in Figure 6c. This result arises from the conservation laws
of energy and angular momentum and the behavior of ${F_{DL} }
\mathord{\left/ {\vphantom {{F_{DL} } {F_0 }}} \right.
\kern-\nulldelimiterspace} {F_0 }$ near the boundary at $r_{_{S}}
$.


\section{CURRENT DENSITIES FLOWING FROM MAGNETOSPHERE INTO
HORIZON AND DISK}

\quad\quad In W03b we calculated the current density flowing from
the BH magnetosphere into the MC region on the horizon in the
state of CEBZMC by using the circuit I. In this paper the three
energy mechanisms of the magnetic extraction are described by
using the circuits I and II, based on which the current densities
flowing from the BH magnetosphere into the horizon and the disk
can be calculated and compared. As argued in W03b the poloidal
current flowing in each loop of circuit I can be written as


\begin{equation}
\label{eq56} I_{BZ}^H \left( {a_ * ,\theta } \right) = I_0
\frac{a_ * \left( {1 - k} \right)}{2\csc ^2\theta - \left( {1 - q}
\right)}, \quad 0 < \theta < \theta _S ,
\end{equation}


\begin{equation}
\label{eq57} I_{MC}^H \left( {a_ * ,\theta ,n} \right) = I_0
\frac{a_ * \left( {1 - \beta } \right)}{2\csc ^2\theta - \left( {1
- q} \right)}, \quad \theta _S < \theta < \theta _L ,
\end{equation}


\begin{figure}
\vspace{0.5cm}
\begin{center}
{\includegraphics[width=6cm]{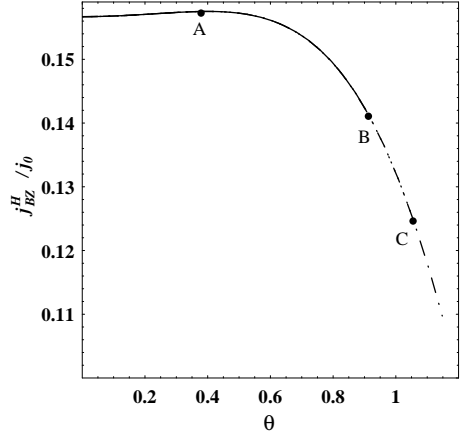}
 \centerline{\quad\quad\quad(a)}
 \includegraphics[width=6cm]{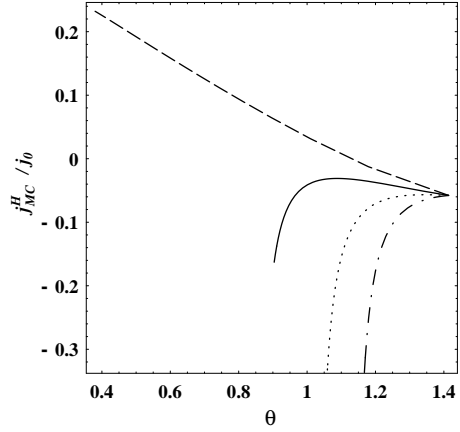}
 \centerline{\quad\quad\quad(b)}}
 \caption{ Current densities versus $\theta $
with $a_ * = 0.9$, $\theta _L = 0.45\pi $, (a) $j_{BZ}^H /
j_{_{0}} $ for $0 < \theta < \theta _S $, where A, B, C are the
end points of the curves corresponding to $n$=4, 5.5, 7,
respectively. (b) $j_{MC}^H  / j_{_{0}} $ for $\theta _S < \theta
< \theta _L $ for $n$ = 4, 5.5, 7 and 9 in dashed, solid, dotted
and dot-dashed lines, respectively.}\label{fig7}
\end{center}
\end{figure}

\begin{figure}
\vspace{0.5cm}
\begin{center}
{\includegraphics[width=6cm]{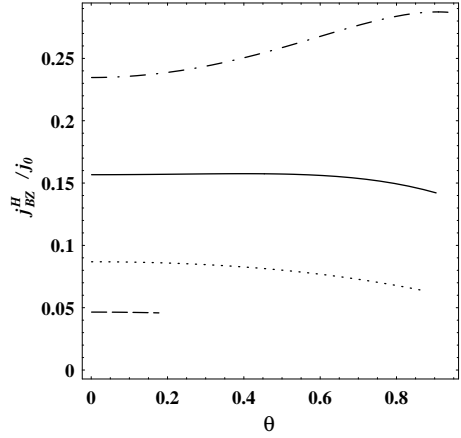}
 \centerline{\quad\quad\quad(a)}
 \includegraphics[width=6cm]{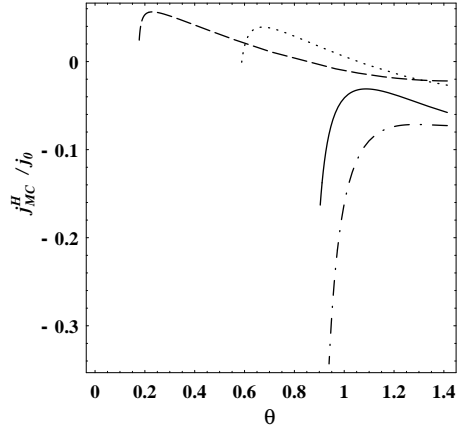}
 \centerline{\quad\quad\quad(b)}}
 \caption{ Current densities versus $\theta$ with $n = 5.5$,
$\theta_L = 0.45 \pi$, (a) $j_{BZ}^H / j_{_{0}}$, (b) $j_{MC}^H /
j_{_{0}}$ for $a_*$ =0.36 , 0.62, 0.9 and 0.998 in dashed, dotted,
solid, and dot-dashed lines, respectively.}\label{fig8}
\end{center}
\end{figure}


\begin{figure}
\vspace{0.5cm}
\begin{center}
{\includegraphics[width=6cm]{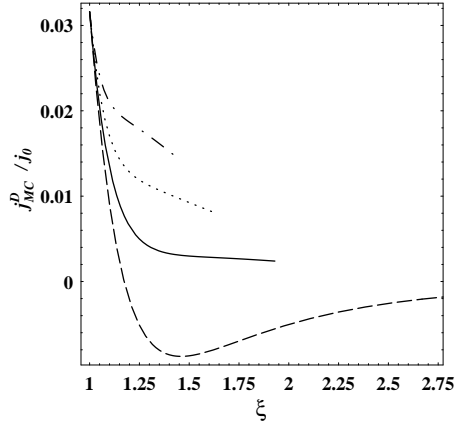}
 \centerline{\quad\quad(a)}
 \includegraphics[width=6cm]{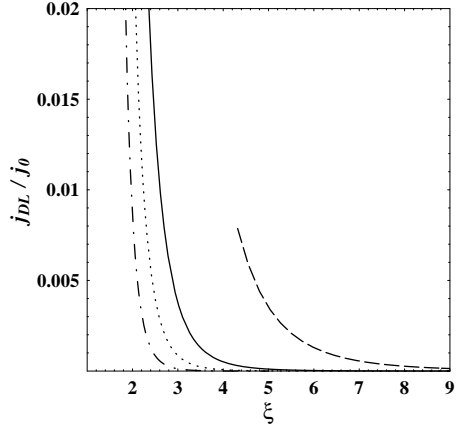}
 \centerline{\quad\quad\quad\quad\quad(b)}}
 \caption{ Current densities versus $\xi$ with $a_* = 0.9$,
$\theta_L = 0.45 \pi$, (a) $j_{MC}^D / j_{_{0}}$, (b)  $j_{_{DL}}
/ j_{_{0}}$ for $n$ =4, 5.5, 7 and 9 in dashed, solid, dotted and
dot-dashed lines, respectively.}\label{fig9}
\end{center}
\end{figure}


\begin{figure}
\vspace{0.5cm}
\begin{center}
{\includegraphics[width=6cm]{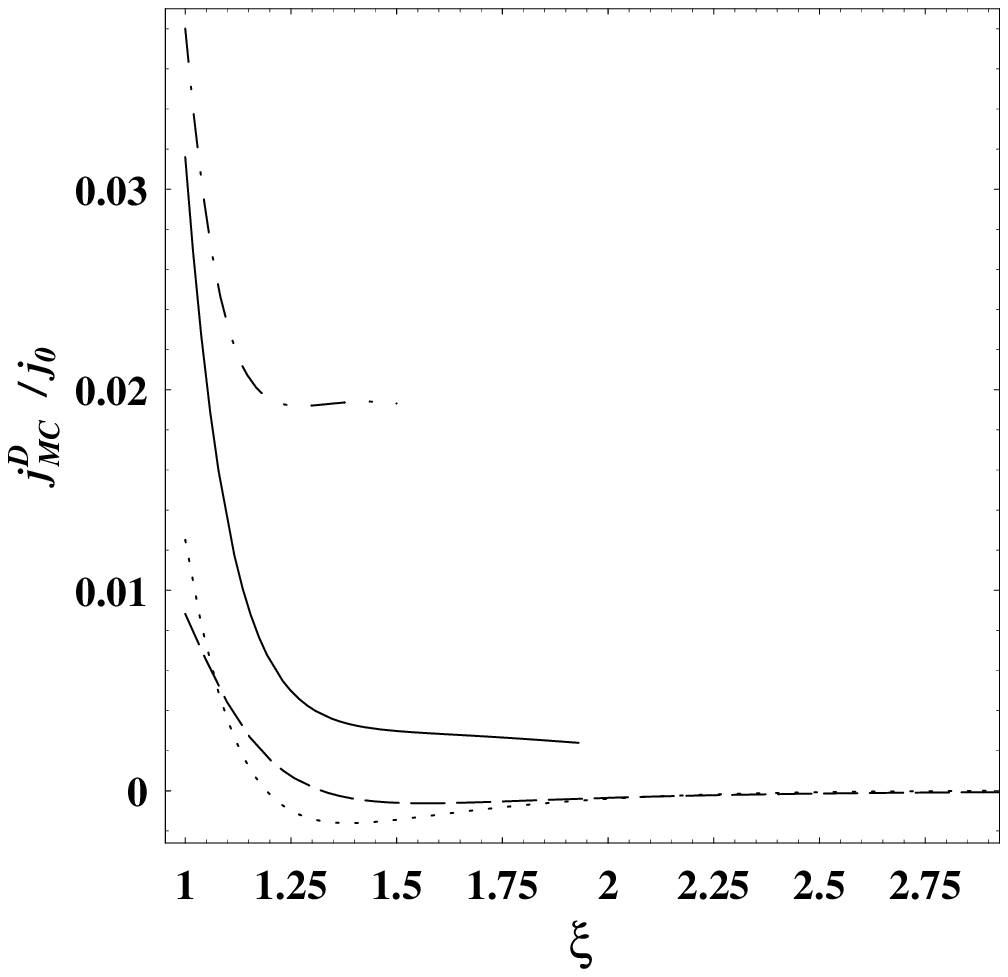}
 \centerline{\quad\quad\quad(a)}
 \includegraphics[width=6cm]{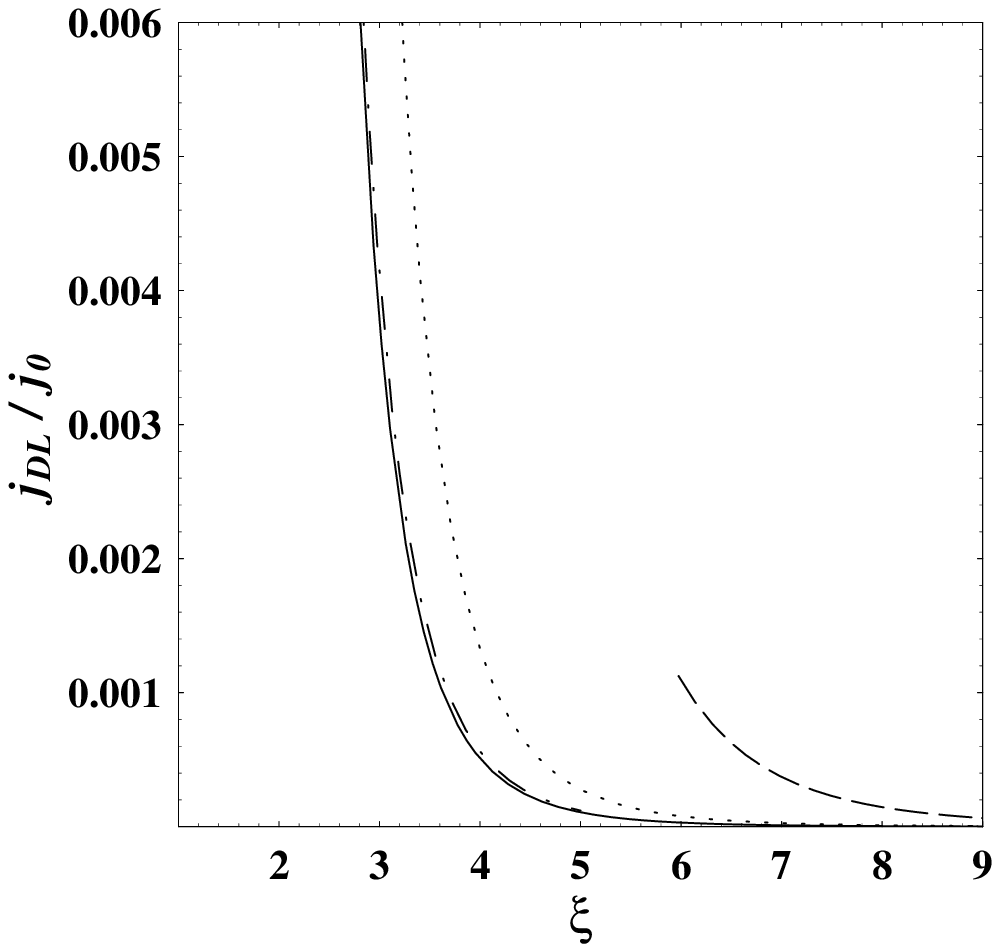}
 \centerline{\quad\quad\quad\quad(b)}}
 \caption{ Current densities versus $\xi$ with $n = 5.5$, $\theta_L = 0.45
\pi$, (a) $j_{MC}^D / j_{_{0}}$, (b) $j_{_{DL}} / j_{_{0}}$ for
$a_*$ = 0.36 , 0.62, 0.9 and 0.998 in dashed, dotted, solid and
dot-dashed lines, respectively.}\label{fig10}
\end{center}
\end{figure}

\noindent where $I_0 = B_H^p M \approx 1.48\times 10^{10}B_4
\left( {M \mathord{\left/ {\vphantom {M {M_ \odot }}} \right.
\kern-\nulldelimiterspace} {M_ \odot }} \right)A$. The currents
$I_{BZ}^H $ and $I_{MC}^H $ are respectively the poloidal currents
flowing on the BZ and MC regions of the horizon, which depend on
the angular coordinate $\theta $ for the given values of $a_ * $
and $n$. Inspecting equations (\ref{eq56}) and (\ref{eq57}), we
find that $I_{BZ}^H $ are generally not equal to $I_{MC}^H $ for
$\beta \ne k$. The conservation of the current at $\theta _S $ is
guaranteed by the current flowing between the magnetosphere and
the horizon. The continuity of the current at the boundary between
the BZ and MC regions of the horizon is discussed in W03b.

\quad\quad Based on the conservation of current, the current
densities flowing from the BH magnetosphere into the above regions
of the horizon are expressed by


\begin{equation}
\label{eq58}\begin{array}{l} j_{BZ}^H \left( {a_ * ,\theta }
\right) = \frac{1}{2\pi \left( {\varpi \rho } \right)_{r =
r_{_{H}} } }\frac{dI_{BZ}^H }{d\theta }\\ \quad\quad = j_{_{0}}
\frac{4\left( {1 - k} \right)M\Omega _H \cos \theta }{\left[ {2 -
\sin ^2\theta \left( {1 - q} \right)} \right]^2}, \quad\quad 0 <
\theta < \theta _S ,
\end{array}
\end{equation}


\begin{equation}
\label{eq59}\begin{array}{l}j_{MC}^H \left( {a_ * ,\theta ,n}
\right) = \frac{1}{2\pi \left( {\varpi \rho } \right)_{r =
r_{_{H}} } }\frac{dI_{MC}^H }{d\theta } \\ \quad\quad= j_{_{0}}
\frac{M\Omega _H }{2 - \sin ^2\theta \left( {1 - q} \right)}\left[
{\frac{4\left( {1 - \beta } \right)\cos \theta }{2 - \sin ^2\theta
\left( {1 - q} \right)} - \sin \theta \frac{d\beta }{d\theta }}
\right],\quad \mbox{ }\theta _S < \theta < \theta _L , \end{array}
\end{equation}

\noindent where $j_{_{0}} \equiv {B_H^p } \mathord{\left/
{\vphantom {{B_H^p } {\left( {2\pi M} \right)}}} \right.
\kern-\nulldelimiterspace} {\left( {2\pi M} \right)} = 0.108\times
B_4 \left( {M \mathord{\left/ {\vphantom {M {M_ \odot }}} \right.
\kern-\nulldelimiterspace} {M_ \odot }} \right)^{ - 1}A \cdot cm^{
- 2}$. For the given value of $a_ * $ and $n$, both $j_{BZ}^H $
and $j_{MC}^H $ vary with angle $\theta $ in the range $0 < \theta
< \theta _S $ and $\theta _S < \theta < \theta _L $, respectively.

\quad\quad According to circuit I the poloidal current flowing on
the MC region of the disk is equal to that flowing on the MC
region of the horizon, i.e.,


\begin{equation}
\label{eq60} I_{MC}^D \left( {a_ * ,\xi ,n} \right) = I_{MC}^H =
I_0 \frac{a_ * \left( {1 - \beta } \right)}{2\csc ^2\theta -
\left( {1 - q} \right)}, \quad 1 < \xi < \xi _S .
\end{equation}

Therefore the current density flowing from the BH magnetosphere
into the disk is


\begin{equation}
\label{eq61} j_{MC}^D \left( {a_ * ,\xi ,n} \right) =
\frac{1}{2\pi \left( {{\varpi \rho } \mathord{\left/ {\vphantom
{{\varpi \rho } {\sqrt \Delta }}} \right.
\kern-\nulldelimiterspace} {\sqrt \Delta }} \right)_{\theta = \pi
\mathord{\left/ {\vphantom {\pi 2}} \right.
\kern-\nulldelimiterspace} 2} }\frac{dI_{MC}^D }{dr}, \quad 1 <
\xi < \xi _S .
\end{equation}

\quad\quad Substituting equation (\ref{eq60}) into equation
(\ref{eq61}), we have the current density flowing from the
magnetosphere into the MC region of the disk as follows,


\begin{equation}
\label{eq62}\begin{array}{l}
 j_{MC}^D \left( {a_ * ,\xi ,n} \right) = \frac{j_{_{0}} a_ * C\left( {a_ * ,\xi } \right)}{2\csc ^2\theta - \left( {1 -
q} \right)}\left[ {\frac{4\left( {1 - \beta } \right)\csc ^2\theta
\cot \theta }{2\csc ^2\theta - \left( {1 - q}
\right)}\frac{d\theta }{d\xi } - \frac{d\beta }{d\xi }}
\right],\mbox{ }1 < \xi < \xi _S , \end{array}
\end{equation}

\noindent where

\begin{equation}
\label{eq63} C\left( {a_ * ,\xi } \right) \equiv \frac{\sqrt {1 +
\xi ^{ - 2}\chi _{ms}^{ - 4} a_ * ^2 - 2\xi ^{ - 1}\chi _{ms}^{ -
2} } }{\xi \chi _{ms}^4 \sqrt {1 + \xi ^{ - 2}\chi _{ms}^{ - 4} a_
* ^2 + 2\xi ^{ - 3}\chi _{ms}^{ - 6} a_ * ^2 } }.
\end{equation}

\quad\quad Applying the same procedure to circuit II, we can
derive the current flowing on the DL region of the disk and the
current density flowing from the BH magnetosphere into the disk as
follows,


\begin{equation}
\label{eq64} I_{DL} = - I_0 \frac{\left( {1 + q} \right)D\left(
{a_ * ,\xi,n } \right)}{2}, \quad \xi _S < \xi < \infty
\end{equation}


\begin{equation}
\label{eq65}\begin{array}{l}
 j_{_{DL}} \left( {a_ * ,\xi ,n} \right) = \frac{1}{2\pi \left(
{{\varpi \rho } \mathord{\left/ {\vphantom {{\varpi \rho } {\sqrt
\Delta }}} \right. \kern-\nulldelimiterspace} {\sqrt \Delta }}
\right)_{\theta = \pi \mathord{\left/ {\vphantom {\pi 2}} \right.
\kern-\nulldelimiterspace} 2} }\frac{dI_{DL} }{dr} \\\\ \quad\quad
= - \frac{j_{_{0}} \left( {1 + q} \right)C\left( {a_ * ,\xi }
\right)}{2}\frac{\partial D\left( {a_ * ,\xi ,n} \right)}{\partial
\xi },\quad \mbox{ }\xi _S < \xi < \infty ,
\end{array}
\end{equation}

\noindent where $C\left( {a_ * ,\xi } \right)$ is expressed by
equation (\ref{eq63}) and $D\left( {a_ * ,\xi } \right)$ is
expressed by

\begin{equation}
\label{eq66}\begin{array}{l} D\left( {a_ * ,\xi ,n} \right) =
\frac{\xi ^{ - n + 2}\chi _{ms}^2 \sqrt {1 + \xi ^{ - 2}\chi
_{ms}^{ - 4} a_ * ^2 + 2\xi ^{ - 3}\chi _{ms}^{ - 6} a_ * ^2 }
}{\xi _S^{n + 1} \left( {\xi ^{3 \mathord{\left/ {\vphantom {3 2}}
\right. \kern-\nulldelimiterspace} 2}\chi _{ms}^3 + a_ * }
\right)\sqrt {1 + \xi ^{ - 2}\chi _{ms}^{ - 4} a_ * ^2 - 2\xi ^{ -
1}\chi _{ms}^{ - 2} } }
\end{array}
\end{equation}

\quad\quad By using equations (\ref{eq58}) and (\ref{eq59}) we
have the current densities $j_{BZ}^H $ and $j_{MC}^H $ varying
with the angle $\theta $ on the horizon for $a_ * = 0.9$ and the
different values of $n$, and those for $n = 5.5$ and the different
values of $a_ * $ as shown in Figures 7 and 8, respectively.

\quad\quad By using equations (\ref{eq62}) and (\ref{eq65}) we
have the current densities $j_{MC}^D $ and $j_{DL}^D $ varying
with the radial parameter $\xi $ on the disk for $a_
* = 0.9$ and the different values of $n$, and those for $n = 5.5$ and the
different values of $a_ * $ as shown in Figures 9 and 10,
respectively.

\quad\quad From Figures 7---10, we find that the distribution
features of the above current densities depend on both the
power-law index $n$ and the BH spin $a_ *$. The current densities
$j_{BZ}^H $ and $j_{_{DL}}$  are always positive, while $j_{MC}^H
$ and $j_{MC}^D $ might change their signs in some cases, which
correspond to the corotation magnetic surface (CRMS) as defined in
W03b.


\section{DISCUSSION}

\quad\quad In this paper we discuss and compare the three powers
of extracting energy magnetically by considering the restriction
of the screw instability to the configuration of the magnetic
field in BH magnetosphere. These powers are derived by using
circuit I for the BZ and MC processes and circuit II for the DL
process. It turns out that the DL power is generally less than the
BZ and MC powers, and it is comparable with the two powers for the
BH spin $a_* $ approaching unity. By using the conservation laws
of energy and angular momentum we discuss the radiation flux from
a quasi-steady thin disk around a Kerr BH. Although the situation
at the boundary between the MC and DL regions is not very
transparent, we obtain the solution for the radiation by resolving
the equation from the conservation laws. By using circuits I and
II we discuss the distribution of the currents and the current
densities on the BH horizon and on the disk for the three energy
mechanisms.

\quad\quad Recently Li (2000b) presented a model for extracting
clean energy from a Kerr BH surrounded by a dense plasma torus,
and argued that this model may be relevant to GRBs, provided that
the magnetic field is strong enough. Li argued that the baryonic
contamination from the plasma in the torus is greatly suppressed
by the magnetic confinement. In fact, the problem of the baryonic
contamination can be also suppressed by the closed field lines in
the MC region in our model, provided that the magnetic field is
strong enough. Inspecting equations (\ref{eq12}) and (\ref{eq19}),
and Figure 4, we find that the maximum of the BZ power could be in
the following range,


\begin{equation}
\label{eq67} 0.05P_0 < P_{BZ}^{\max } < 0.1P_0 ,
\end{equation}

\noindent which implies that $P_{BZ}^{\max } $ could attains $
\sim 10^{51}erg \cdot s^{ - 1}$ for $M = 7M_ \odot $ and $B_4 =
10^{11}$. Although $P_{BZ}^{\max } $ seems much more than needed
for powering the beamed GRBs, we can adjust the value of $P_{BZ} $
to fit the beamed GRBs by taking the adequate values of $M$ and
$B_4 $ in our model. In addition, the DL process discussed in this
paper could be another energy mechanism for extracting clean
energy to GRBs. We shall apply this model to GRBs and present a
detailed calculation in the future work.\\

\noindent\textbf{Acknowledgments. }This work is supported by the
National Natural Science Foundation of China under Grant Numbers
10173004, 10373006 and 10121503.

{}

\end{document}